\documentclass[
    ,final            
  ]
  {aipproc}

\layoutstyle{6x9}

\begin{document}

\title{The Advanced X-ray Timing Array (AXTAR)}


\author{Deepto Chakrabarty}{
  address={MIT Kavli Institute for Astrophysics and Space Research, 
Cambridge, MA 02139, USA}
}

\author{Paul S. Ray}{
  address={Space Science Division, Naval Research Laboratory, Washington,
DC 20375, USA}
}

\author{Tod E. Strohmayer}{
  address={Astrophysics Science Division, NASA Goddard Space Flight Center, 
Greenbelt, MD 20771, USA}
}

\author{the~AXTAR~Collaboration}{
  address={MIT Kavli Institute for Astrophysics and Space Research, 
Cambridge, MA 02139, USA}
}

\begin{abstract}
  AXTAR is an X-ray observatory mission concept, currently under study
  in the U.S., that combines very large collecting area, broadband
  spectral coverage, high time resolution, highly flexible scheduling,
  and an ability to respond promptly to time-critical targets of
  opportunity.  It is optimized for submillisecond timing of bright
  Galactic X-ray sources in order to study phenomena at the natural
  time scales of neutron star surfaces and black hole event horizons,
  thus probing the physics of ultradense matter, strongly curved
  spacetimes, and intense magnetic fields.  AXTAR's main instrument is
  a collimated, thick Si pixel detector with 2--50~keV coverage and
  8~m$^2$ collecting area.  For timing observations of accreting
  neutron stars and black holes, AXTAR provides at least an order of
  magnitude improvement in sensitivity over both RXTE and
  Constellation-X.  AXTAR also carries a sensitive sky monitor that
  acts as a trigger for pointed observations of X-ray transients and
  also provides continuous monitoring of the X-ray sky with 20$\times$
  the sensitivity of the RXTE ASM. AXTAR builds on detector and
  electronics technology previously developed for other applications
  and thus combines high technical readiness and well understood cost.
\end{abstract}

\maketitle


\section{INTRODUCTION}

The natural time scales near neutron stars (NSs) and stellar mass
black holes (BHs) are in the millisecond range.  These time scales
characterize the fundamental physical properties of compact objects:
mass, radius, and angular momentum. For example, the maximum spin rate
of a NS is set by the equation of state of the ultradense matter in
its interior, a fundamental property of matter that still eludes us.
Similarly, orbital periods at a given radius near a BH are set by the
BH's mass, angular momentum, and the laws of relativistic gravity.
Although it was recognized for decades that the measurement of such
time scales would provide unique insight into these compact stars and
their extreme physics, it was not until the 1995 launch of the Rossi
X-ray Timing Explorer (RXTE; \citep{brs93}) that oscillations on these
time scales were actually detected from accreting NSs and stellar-mass
BHs.  This conference celebrates RXTE's discovery of millisecond
oscillations that trace the spin rate of accreting NSs.  RXTE has also
discovered millisecond oscillations from accreting BHs with
frequencies that scale inversely with BH mass and are consistent with
the orbital time scale of matter moving in the strongly curved
spacetime near the BH event horizon.

While RXTE revealed the existence of these phenomena, it lacks the
sensitivity to fully exploit them in determining the fundamental
properties of NSs and BHs.  Here, we describe the Advanced X-ray Timing
Array (AXTAR), a new mission concept boasting an order of magnitude
improvement in sensitivity over RXTE\footnote{A similar
  mission concept was previously discussed by Phil Kaaret and
  collaborators \citep{kgl+01,kaa04}.}.  AXTAR was initially proposed
as a medium-class probe concept for NASA's 2007 Astrophysics Strategic
Mission Concept Studies.  A modified version of AXTAR is also being
studied as a NASA MIDEX-class mission concept.

\begin{table}
\begin{tabular}{p{1.5in}p{3in}}
\hline
\tablehead{1}{c}{b}{Science Objective} 
  & \tablehead{1}{c}{b}{AXTAR Observations} \\
\hline
NS mass, radius, EOS  & X-ray burst oscillations. kHz QPOs. Accreting
                        ms pulsars. Asteroseismology with 
                        magnetar oscillations. \\ \hline
Strongly curved spacetimes
                      & BH oscillations. Broad Fe lines. Phase-resolved
                        spectroscopy of low-freq QPOs. AGN
                        monitoring. \\ \hline
Physics of nuclear burning
                      & Thermonuclear X-ray bursts and superbursts. \\ \hline
Physics of accretion  & Accreting msec pulsars. kHz QPOs in NSs. \\ \hline
Physics of jets       & NS and BH transients.  \\ \hline
Physics of mass transfer
                      & X-ray pulsar orbital evolution.  \\ \hline
Multipolar magnetic field components of pulsars
                      & Hard X-ray cyclotron lines in high-mass binaries.
                        Magnetar pulse profiles. \\ \hline
\end{tabular}
\caption{Selected AXTAR Science Topics}
\label{tab:a}
\end{table}

\section{SCIENCE OBJECTIVES}

AXTAR will study a broad range of topics in the astrophysics of NSs
and BHs as well as the physics of ultradense matter, strongly curved
spacetime, and intense magnetic fields (see Table~1).  The mission
design was driven by the requirements for three key efforts:

{\bf Neutron Star Radii and the Equation of State of Ultradense
  Matter.} AXTAR will make the first precise measurements of NS radii,
using pulse profile fitting of millisecond brightness oscillations
near the onset of thermonuclear X-ray bursts.  Gravitational
self-lensing by the NS suppresses the amplitude of observed
oscillations by allowing an observer to ``see'' more than just the
facing hemisphere, an effect set by the ratio of mass to radius, $M/R$
\citep{ml98,pfc83,szs+98}.  Additionally, the pulse shape of the
oscillations is influenced by the NS rotational velocity, $v_{\rm
  rot}\propto R\,\nu_{\rm spin}$, where the NS spin frequency
$\nu_{\rm spin}$ is known from the oscillations
\citep{brr00,wml01,moc02}.  The pulsed surface emission encodes
information about the NS structure, and modeling of high
signal-to-noise profiles can be used to constrain $M$ and $R$
separately \citep{str04}.  Precise measurement of NS radii at the 5\%
level is the key to determining the equation of state of ultradense
matter \citep{lp01}, one of the most fundamental questions arising at
the interface of physics and astrophysics.

{\bf Black hole oscillations and the physics of extreme gravity.}
AXTAR will determine if high-frequency oscillations from BHs are a
direct tracer of mass and spin.  In several systems, RXTE has detected
high-frequency QPO pairs whose properties suggest such a relationship
\citep{rm06}. By extending the detection threshold down a 20$\times$
to $\sim$0.05\% amplitude, AXTAR will test this prospect in
two ways: (1) by detecting additional examples of rapid QPO pairs in
other BHs to test the $1/M$ scaling of the observed frequencies with a
larger source sample; and (2) by detecting additional weaker QPO modes
in the existing systems, as predicted by models for the BH
oscillations [see, e.g., 13-17].

{\bf Continuous long-term monitoring of the variable X-ray sky.}  The
AXTAR Sky Monitor will {\em continuously} monitor hundreds of X-ray
sources in addition to serving as a trigger for target-of-opportunity
observations of active X-ray transients. A sensitive sky monitor
with nearly continuous all-sky coverage can serve as a {\em
  primary} science instrument for a wide variety of investigations
using an all-sky sample, including: 
daily flux and spin monitoring of accretion-powered pulsars,
phase-coherent tracking of the spin-down of magnetars and detection of
glitches, long-term monitoring of bright AGNs, superbursts, X-ray
flashes, etc. 

\begin{figure}
  \includegraphics[height=.23\textheight]{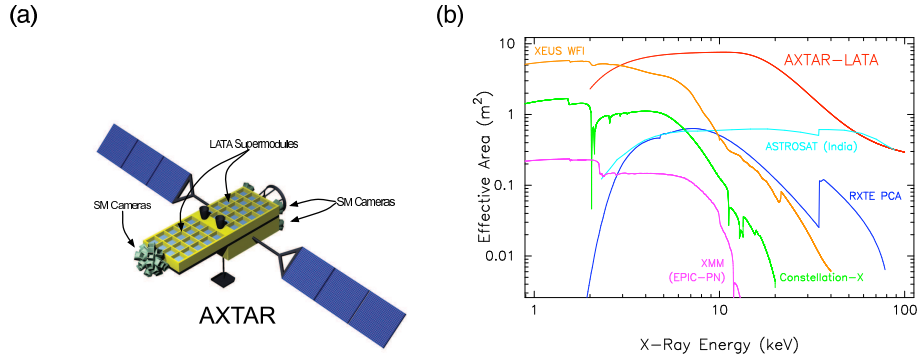}
  \caption{(a) A notional depiction of the AXTAR mission concept. 
    (b) The effective area curves for AXTAR and several other missions
    with X-ray timing capabilities.}
\end{figure}

\section{MISSION CONCEPT}

The AXTAR mission concept assumes a 3-axis stabilized spacecraft in
low-Earth orbit.  A notional illustration of the spacecraft layout is
shown in Figure~1a, and the mission parameters are summarized in
Table~2.  Like RXTE, AXTAR is optimized for study of X-ray transients
anywhere in the sky by combining a small solar avoidance angle with
the ability to repoint promptly ($<$1~d) to targets of opportunity (TOOs). 


\begin{table}
\begin{tabular}{p{1.5in}p{1in}p{1.2in}}
\hline
\tablehead{1}{c}{b}{Parameter} 
  & \tablehead{1}{c}{b}{Baseline}
  & \tablehead{1}{c}{b}{Study Range} \\
\hline
\multicolumn{3}{c}{LARGE AREA TIMING ARRAY (LATA)} \\
Effective area         & 7.6 m$^2$      & 6--10 m$^2$ \\
Low E threshold        & 2 keV          & 1--3 keV  \\
High E threshold       & 50 keV         & 25--80 keV \\
Energy resolution      & 600 eV         & 300--900 eV \\
Deadtime               & 10\% on 10 Crab& 1--20\% on 10 Crab \\
Time resolution        & 1 $\mu$s       & 1--100 $\mu$s \\ \hline

\multicolumn{3}{c}{SKY MONITOR (SM)} \\
Sensitivity (1 day)    & 5 mCrab        & 1--25 mCrab \\
Sky coverage           & 4$\pi$ sr     & $>$1 sr  \\
Source loc. acc.       & 1 arcmin       & 0.5--5 arcmin \\ \hline

\multicolumn{3}{c}{AXTAR MISSION} \\
Solar avoidance angle  & 30$^\circ$     & 20$^\circ$--50$^\circ$ \\
Telemetry rate         & 2 Mbps         & 1--10 Mbps \\
Slew rate              & 10$^\circ$/min & 5$^\circ$--25$^\circ$/min\\ \hline

\end{tabular}
\caption{AXTAR Mission Parameters}
\label{tab:b}
\end{table}

AXTAR's principal instrument is the Large Area Timing Array (LATA), a
collimated (1$^\circ$ FWHM), thick (1~mm) Si pixel detector with
2--50~keV coverage and 8~m$^2$ collecting area.   Pixelated
solid-state detectors constructed out of large Si wafers promise
better performance, higher reliability, and lower cost than gas
proportional counters; they also leverage massive industrial
investment in integrated circuit technology.  LATA consists of 1000 of
the largest detectors that can be made from a single 150~mm diameter
Si wafer, coupled with electronics optimized for extremely low
noise and low power.   The effective area curve for AXTAR-LATA is
compared to several other missions in Figure~1b. An improvement of
over an order of magnitude over RXTE is achieved.

The AXTAR Sky Monitor (SM) comprises a set of 32 modest-size coded
aperture cameras, each consisting of a 2D-pixelated Si detector
(essentially identical to those used in the LATA) with 300~cm$^2$ area
and 2--25~keV coverage.  Each detector will view the sky through a 2-D
coded mask and will have a $40^\circ \times 40^\circ$ (FWHM) field of
view. The cameras will be mounted with pointing directions chosen to
cover 4$\pi$ sr.  The AXTAR-SM will be able to monitor most celestial
locations with a 60--100\% duty cycle (depending on final orbit
choice), as compared to the $\sim$3\% duty cycle of RXTE/ASM.  
It will also have timing capability as well as a 1-day sensitivity of
a few mCrab (20$\times$ better than RXTE/ASM).  This will
allow early detection (and TOO triggering) of much fainter transients
as well as enabling the continuous monitoring and timing programs
mentioned above.



\end{document}